\documentclass[oneside,12pt,doublespacing]{elsart}

\usepackage{graphicx,latexsym,amssymb,wasysym,pifont}

\newcommand{\degr}{\ensuremath{^\circ}~}

\def\prd{Phys.\ Rev.\ D\ }
\def\nat{Nature}

\begin{document}
\begin{frontmatter}

\title{Moon and Sun shadowing effect in the MACRO detector}
\author[12]{M.~Ambrosio}, 
\author[7]{R.~Antolini}, 
\author[13]{A.~Baldini}, 
\author[12]{G.~C.~Barbarino}, 
\author[4]{B.~C.~Barish}, 
\author[6,b]{G.~Battistoni}, 
\author[2]{Y.~Becherini},
\author[1]{R.~Bellotti}, 
\author[13]{C.~Bemporad}, 
\author[10]{P.~Bernardini}, 
% ...
\address[1] {Dipartimento Interateneo di Fisica del Politecnico-Universit\`a  di Bari and INFN, 70126 Bari, 
 Italy}
\address[2] {Dipartimento di Fisica dell'Universit\`a  di Bologna and INFN, 40126 
Bologna, Italy }
% ...
\address[4] {California Institute of Technology, Pasadena, CA 91125, USA }
% ...
\address[6] {Laboratori Nazionali di Frascati dell'INFN, 00044 Frascati (Roma), Italy  }
\address[7] {Laboratori Nazionali del Gran Sasso dell'INFN, 67010 Assergi (L'Aquila), 
 Italy  }
% ...
\address[10] {Dipartimento di Fisica dell'Universit\`a  di Lecce and INFN, 73100 Lecce, 
 Italy  }
% ...
\address[12] {Dipartimento di Fisica dell'Universit\`a  di Napoli and INFN, 80125 
Napoli, Italy  }
\address[13] {Dipartimento di Fisica dell'Universit\`a  di Pisa and INFN, 56010 Pisa, 
Italy  }
% ...
\address[b] {Also INFN Milano, 20133 Milano, Italy  }
% ...
\corauth[ng]{Corresponding author.\\{\it E-mail address:} giglietto@ba.infn.it}
\begin{abstract} 
Using data collected by the MACRO experiment from 1989 to the end of its
operations in 2000, we have studied in the underground muon flux the shadowing
effects due to both the Moon and the Sun. We have observed the shadow cast by 
the Moon at its apparent position with a significance of 6.5 $\sigma$. The 
Moon shadowing effect has been used to verify the pointing capability of the 
detector and to determine the instrument resolution for the search of muon 
excesses from any direction of the celestial sphere. The dependence of the 
effect on the geomagnetic field is clearly shown by splitting the data sample 
in day and night observations. The Sun shadow, observed with a significance 
of 4.6 $\sigma$ is displaced by about 0.6\degr from its apparent position. 
In this case however the explanation resides in the configuration 
of the Solar and Interplanetary Magnetic Fields, which affect the propagation 
of cosmic ray particles between the Sun, and the Earth. The displacement of 
the Sun shadow with respect to the real Sun position has been used to 
establish an upper limit on the antimatter flux in cosmic rays of about 
48\% at 68\% c.l. and primary energies of about 20 TeV.
\end{abstract}

\end{frontmatter}

\section{Introduction} 

MACRO was a large area underground detector located in Hall~B of the Gran 
Sasso National Laboratory (LNGS) in Italy at an average depth of~3700~m.w.e, 
13\degr 34\'~E longitude and 42\degr 27\'~N latitude. A detailed description 
of the detector can be found in \cite{MACRO_DEC}. The experiment was 
primarily designed to search for monopoles and rare particles in the cosmic 
rays, including high energy neutrinos and muons from cosmic point sources 
\cite{MUASTRO_RAP}. These sources can be inferred from an excess of
muons 
 above a nearly isotropic 
background in a particular region of the sky \cite{MUASTRO_PAP,NIASTRO_PAP}. 
An important requirement for any kind of detector using this technique is the 
determination of the best signal/background ratio, which is related to the 
angular resolution. \\ 
It was originally suggested by Clark \cite{clark} that an observed narrow 
angle ``shadow'' in the cosmic ray flux due to the absorption by the Sun and 
the Moon can be useful to test the angular resolution\cite{hegra96} and the pointing 
ability of cosmic ray detectors\cite{MUASTRO_PAP,NIASTRO_PAP}. 
 Moreover, the positive determination of the Moon shadow in the expected position,
validates the analyses of coincident data between different
detectors as MACRO together EAS-TOP\cite{MACRO-EASTOP} 
or other detectors\cite{MACRO-GRACE}.

However, since the angular diameter of these bodies 
is about 0.5\degr wide, only detectors having good angular resolution and 
sufficient statistics have the possibility to detect this signal. 
Several large air shower arrays 
\cite{cygnus,eastop,hegra,tibet93,casa,milagro,clue}
 and shallow depth detectors \cite{L3+CO} have observed these 
effects. Deep underground observations ($>$2000 m.w.e.) 
have greater difficulties in observing the shadows, due to the greatly 
reduced underground muon flux. However MACRO \cite{MACRO_MOON}, SOUDAN2 
\cite{soudan2} and LVD \cite{lvd} have all collected data continuously for about 
10 years, thereby gaining the sufficient sensitivity to observe the
effect. 

Underground muons are the remnants of the air showers initiated by the 
collisions of primary cosmic rays with air nuclei. The secondary muons which
 reach the MACRO detector had to cross a minimum rock overburden of 
3200 hg~cm$^{-2}$. This corresponds to a minimum 
energy of the muon at the surface of about 1.4 TeV and to a primary proton 
with median energy around 22 TeV. Since the Earth Magnetic field is 
approximately (to within 10\%) a dipole field, and its  strength is 0.5~G 
at the surface, a single charged particle crossing it will acquire a 
transverse momentum of about 25 GeV/c for a path integral over few 
Earth' radii.  The average displacement of muon trajectories 
due to this effect,as viewed in the detector, will be 
$[0.15\degr Z/ (E_{p}/10~TeV)]$, eastward for positive primaries,
\cite{urban,heintze}.
Therefore the apparent position of the Moon shadow will appear moved to the west direction (with respect 
to the Earth-Moon direction) by the same amount.

The motion in time of the Earth's magnetic pole around its average
position,
introduces a smearing of the 
shadow when the effect is observed over many years \cite{artemis}. 
Additional smearing can be produced by the fact that the observations are 
made at different angles and at different times of the year.

The Sun shadowing effect is more complicated to estimate. Particles
shadowed by the Sun are traveling in the direction of the
Sun-Earth axis and therefore are traveling through the solar magnetic
field and the Interplanetary Magnetic Field (IMF).  The IMF is due to
the electric currents in the Sun that generate a complex magnetic
field which extends out into the interplanetary space. As the Sun's
magnetic field is carried out through the solar system by the solar
wind and the Sun is rotating, the rotation winds up the magnetic field
into a large rotating spiral, known as the Parker spiral.  The
magnetic field is primarily directed outward from the Sun in one of
its hemispheres, and inward in the other.  This causes opposite
magnetic field directions in the Parker spiral.  The thin layer
between the different field directions is described as the neutral
current sheet. Since this dividing line between the outward and inward
field directions is not exactly on the solar equator, the rotation of
the Sun causes the current sheet to become "wavy", and this waviness
is carried out into interplanetary space by the solar wind.  For this
reason the IMF shows a sector structure with field directions
reversing across the sector boundaries \cite{wilcox,svalga};
therefore, in some sectors the magnetic field points inward, and
outward in others. Moreover, this structure varies with the solar
activity cycle; a complete simulation  must
thus take into account the solar activity phase and the sectors
encountered by the particle traveling
in our solar system\cite{jap,cin,suga}.

The possibility to use the Moon and Sun shadows as mass spectrometers was 
first explored by Lloyd-Evans \cite{Lloyd-Evans}; following this idea 
Urban et al. \cite{urban} proposed this method as a way to search for 
antimatter in primary cosmic rays. If there is a significant antimatter
component in the primary CRs in the TeV energy region, then the magnetic 
fields should deflect them in opposite direction with respect to the matter 
component; therefore the proton component should be deflected by the 
geomagnetic field eastward and the antiproton component westward. As a 
consequence the shadow produced by the proton component should be to the west 
with respect to the Moon center, while the resulting shadow due to antiprotons
should be to the east. However to resolve the images, the two disks must be 
far from each other, at least by the disk diameter itself.

The shadow of the Moon was previously observed by MACRO using a partial data
sample\cite{MACRO_MOON}. 
In this paper we present the measurement of the Moon and Sun shadows using 
the full data set of muons collected by the MACRO detector.
 Using the result obtained for the Sun we are able to set an 
upper limit to the antiproton flux of TeV energy at the Earth.

\section{The muon data sample and the expected background}

\subsection{Muon data sample}

The muon sample used for the present analysis includes all events collected
from the start of MACRO data taking in February, 1989 through the end of
2000. The sample totals $50.0 \times 10^6$ events collected over
74,073 hours of livetime. During the first part of this period, the apparatus 
was under construction.  The three main detector configurations
included one ($A_{eff}\Omega \approx 1,010$~m$^2$~sr) and six ($A_{eff}\Omega \approx 5,600$~m$^2$~sr) supermodules without
the attico (the upper part of the detector described in\cite{MACRO_DEC}),
 and finally the full
six supermodules with attico ($A_{eff}\Omega \approx 6,600$~m$^2$~sr).
Approximately 60\% of the data sample was obtained during periods when 
MACRO had full acceptance.

\begin{figure}[t]
\begin{center}
\includegraphics[height=15.cm,width=15.cm]{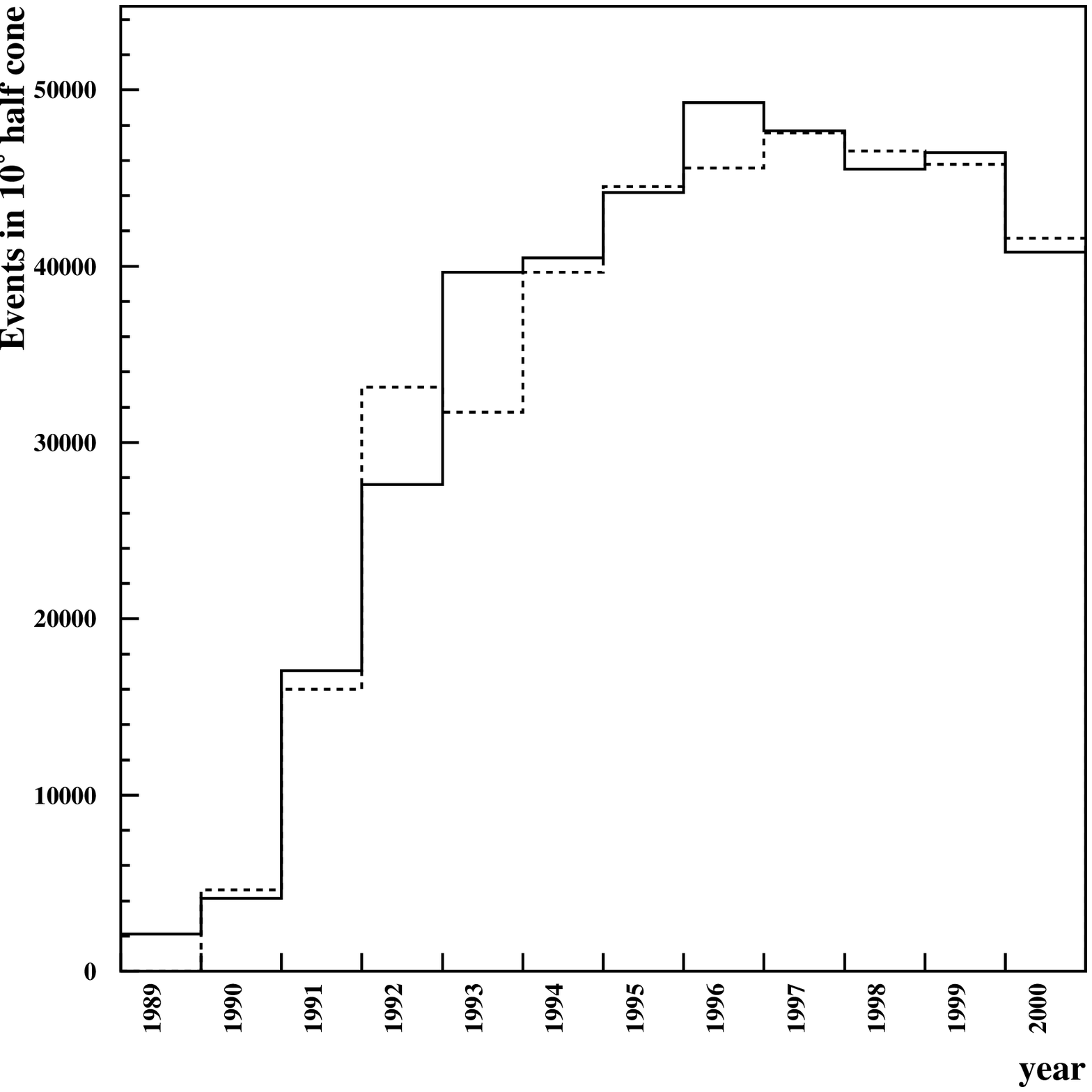}
\caption{Number of events in a 10$^\circ$ half-angle cone 
from the Moon direction (solid histogram) and from the Sun direction (dashed
histogram) collected during each year of MACRO data taking.}
\label{sunvsyear}
\end{center}
\end{figure}

The run and event was used to select good quality events 
were the same as in our first analysis of the Moon shadow
\cite{MACRO_MOON} and for the muon-astronomy searches \cite{MUASTRO_PAP}.
Thus, only muon events contained in a half-angle cone of $10^\circ$ and 
centered on the Moon and the Sun have been retained for further analysis.
The number of  events that passed all cuts in the Moon window was
404988,  almost doubling the 
previous statistics, and 396662 events in the Sun window.
The rate of accumulation of the events in the two windows is given in 
Fig.~\ref{sunvsyear}. As visible in Fig.~\ref{sunvsyear} the effects of
a  reduced detector livetime, due to the
installation of the final configurations in years 1992-1993, produced
a different esposure for the  Sun and The Moon.

We note that during
the period  when MACRO was operative
the solar activity went from its maximum phase in 1991, through a minimum 
phase around 1996/97 to the beginning of the next maximum in 2001.
However almost 62\% of the total events were accumulated from 1994 onward, 
close to the period of minimum solar activity when the neutral sheet was 
probably lying close to the ecliptic plane with little warping \cite{balogh}.
The period  is contained within the so-called A$>$0 phase of the solar cycle 
during which the magnetic field of the Sun is directed outward at the north 
pole of the Sun and inward in the South.  

As in our previous analysis, for each real event observed, we produced 25 
simulated background events, by randomly coupling muon directions and times, 
as explained in \cite{MUASTRO_PAP}. This technique was used to
estimate the expected events from any direction of the sky, including
the Sun and the Moon.

\section{Shadow of the Moon}
The Moon is a celestial body moving quickly in the sky. Thus, to perform the 
analysis it is necessary to make a very precise and accurate computation 
of its position.
The topocentric position of the moon was computed at the arrival time
of each event using the database of ephemerides available
from the Jet Propulsion Laboratory, JPLEPH \cite{jpl}.  A correction for
the parallax due to MACRO's instantaneous position on the Earth was then
applied to each ephemeris position \cite{duffet}.

\subsection{Event deficit around the Moon}

\begin{figure}[t]
\begin{center}
\includegraphics[height=15cm,width=15cm]{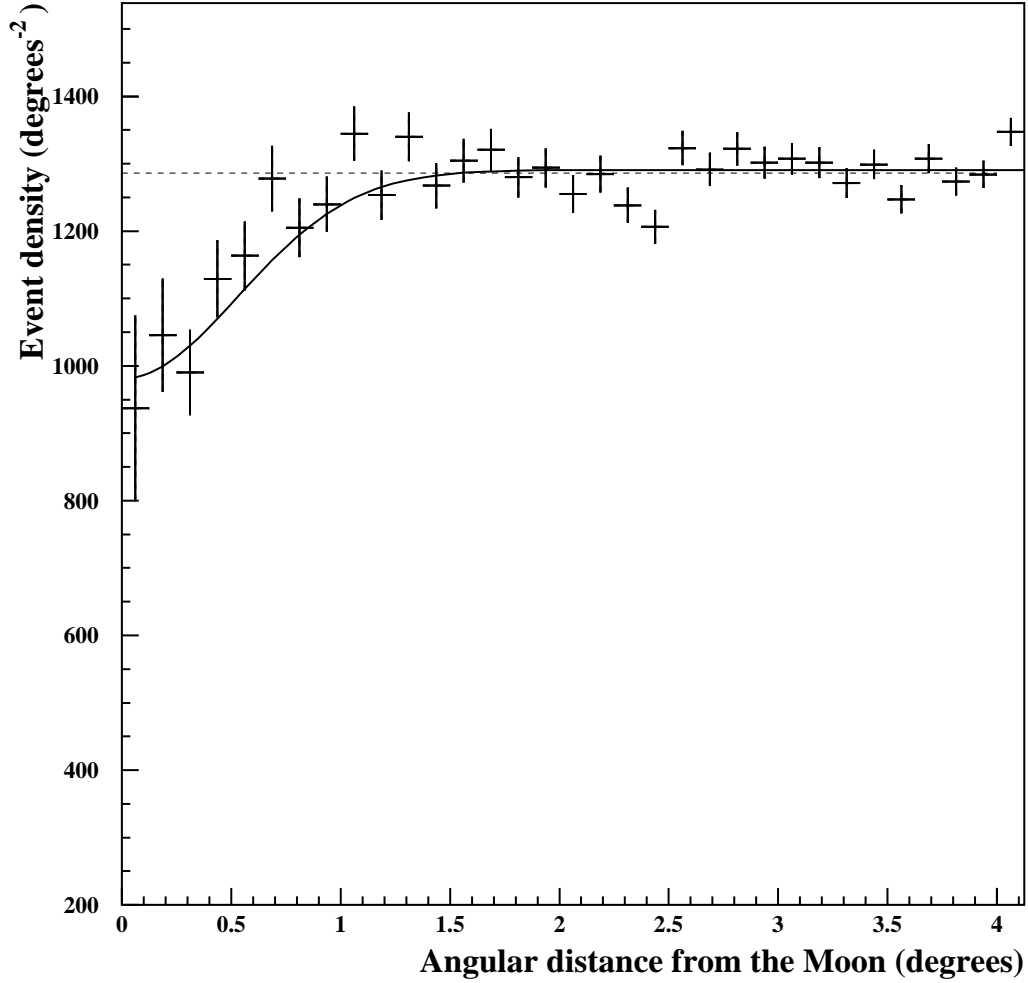}
\caption{Event density vs angular distance from the moon 
  center in bins of equal angular width.  The width of each bin 
is   $0.125^\circ$.  The dashed curve is the average
  expected background computed from 25 background Monte Carlo samples.  
The solid curve shows the expected event density computed for an angular
  resolution of the MACRO apparatus in the hypothesis of a PSF
  normally distributed. The resolution value from the fit is
  $0.55\degr\pm 0.05\degr$.  }
\label{moondens}
\end{center}
\end{figure}

The distribution of events with arrival directions close to the Moon 
direction can be used to make visible the event deficit around the Moon 
disk. The average disk radius is computed for each event of the 404988
events in the sample giving a value of $0.26\pm0.01$. 
 Fig. \ref{moondens} shows the distribution 
$\frac{1}{\pi}\frac{dN}{d\theta^2} \mathrm{vs.} \theta$, 
where $\theta$ is the angular distance between the calculated Moon 
center position in celestial coordinates and the muon arrival direction. 
The significance of the deficit can be calculated by fitting this 
distribution with a function of the form \cite{cygnus,soudan2-00}:\\
\begin{equation}
  \frac{dN_\mu}{d\theta^2}=k
(1-\frac{\theta_M^2}{2\sigma^2}e^{-\frac{\theta^2}{2\sigma^2}})
\end{equation}
where  $\theta_M=0.26^\circ$ is the average value of the angular radius of
the Moon, previously computed, $\sigma$ is the detector angular resolution and k
the event density.   

This function represents the signal produced by the Moon absorption effect 
when the detector point spread function (PSF) can be assimilated to a 
bidimensional normal distribution. We know that in our case this is not 
really appropriate because muons observed in the apparatus undergo multiple 
Coulomb scattering in the overburden rock, which generates long tails in the 
measured angular distributions. Nonetheless, this simple analysis gives 
results consistent with the one presented in the next section where we  
take into account the real PSF of the apparatus by using a maximum likelihood 
method. The fit of the data to the above function has a $\chi^{2}=54.8$ for
31 D.o.F. and a $\chi^{2}=95.9$ for a flat distribution (32 D.o.F.). The 
difference between these values suggests that the chance probability of the 
observed deficit is $\le 10^{-9}$, equivalent to a statistical significance 
of about $6 \sigma$. The fitted values of the parameters are
$\rm k=(1294\pm 5) ~{\rm events/degree^2}$ and $\sigma=(0.55\pm0.05)\degr$.
The value of k obtained by the fit can be used to have a 
rough estimate of the number of missing events (blocked by the Moon disk): 
$\rm\pi k\cdot\theta_M^2 = 275\pm 1$. We also estimated the number of
missing events by considering the number of expected background events 
in the 10\degr cone  windows, N$_{bck}$ = $405100 \pm 127$,  
calculated as mentioned above. Thus, the number of blocked events is 
$405100\cdot (\frac{0.26\degr}{10\degr})^2 =274 \pm 1 $, in agreement with 
the result of the fit. Since the livetime of the apparatus for the entire 
sample is about 74000 hr, the rate of the events obscured by the Moon is 
about 3 events/month.

\begin{figure}[t]
\begin{center}

\includegraphics[height=15cm,width=15cm]{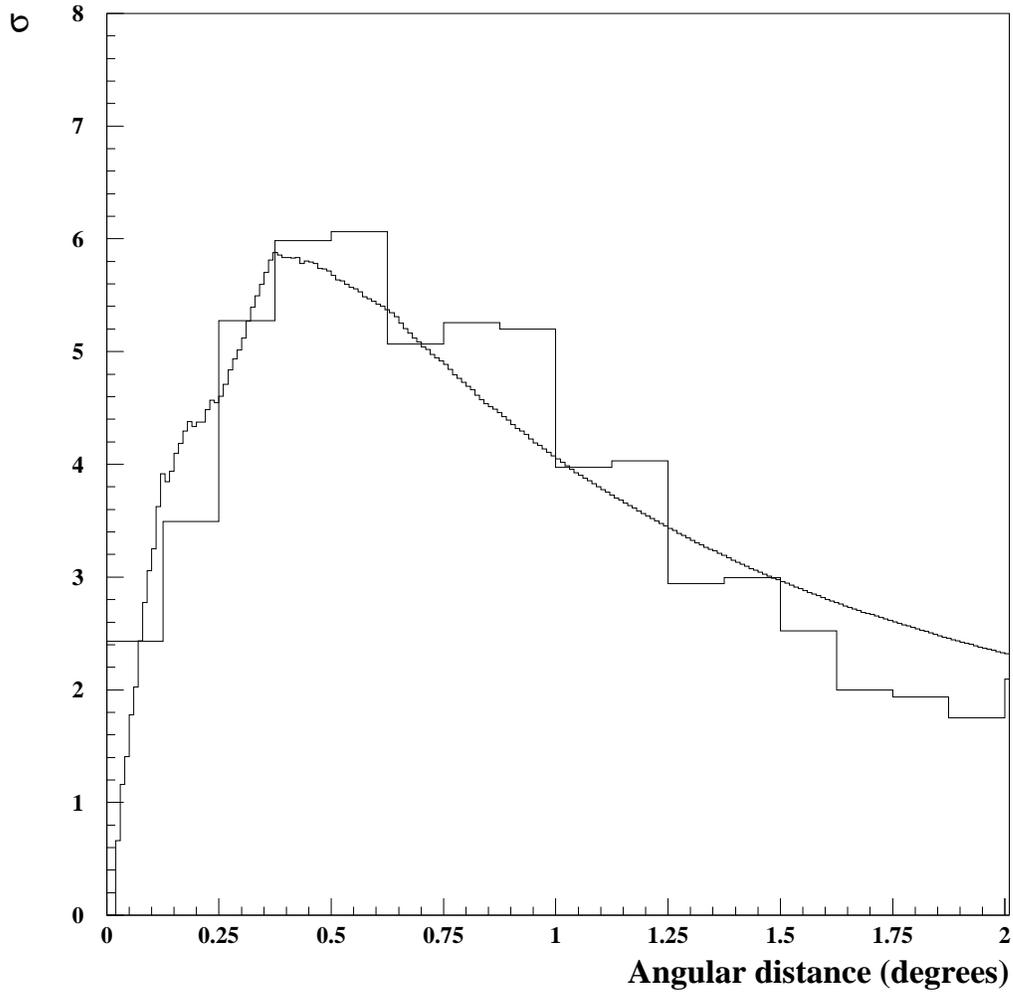}
\caption{Deviation from the expected number of events computed as 
$\sigma=\frac{N_{exp}-N_{obs}}{\sqrt{N_{exp}}}$
  versus the angular distance from the Moon center. Superimposed
  (continuous line) the
  simulated distribution of the deviations expected using the modified
  psf distribution for an extended source of 0.26\degr radius.
}

\label{sigma_int} 
\end{center}
\end{figure}

The observed signal can then be used to optimize the bin size for astronomy 
searches \cite{MUASTRO_PAP}. 
In fact we can choose the size of a circular window centered on the Moon that 
maximizes the statistics $\frac{N_{exp}-N_{obs}}{\sqrt{N_{exp}}}$,
being  $N_{obs}$ integrated the number of events observed in the Moon 
direction and $N_{exp}$ the integrated number of expected events in the same
direction.
The behavior of this quantity versus the angular distance from the Moon 
center, is shown in Fig.\ref{sigma_int}. 
 This figure suggests the choice of bin 0.5\degr radius 
wide as the optimal one for maximizing the signal if there is a source at 
the center of the bin \cite{MUASTRO_PAP}.

\subsection{Maximum likelihood analysis}
In the simple one dimensional analysis above, we have implicitly assumed 
that the position of the moon's shadow is known and the significance 
calculated as the PSF of the apparatus is a perfect two dimensional normal 
distribution. However geomagnetic effects, displacement of the shadow and  the
distortions introduced by the true PSF,
cannot be easily taken into account using this analysis.
 For all these reasons we have developed 
a binned likelihood method, based on a {\it a priori} knowledge of
the MACRO PSF (MPSF), a technique originally developed by COS-B\cite{COS-B}.\\

The MPSF was accurately determined by using double muon events
in the detector. Muon pair events are produced in the decays of pions
and kaons produced in the primary cosmic ray
interactions; these muons therefore come from about 20~km above the
apparatus and have  small initial
separation angles; thus they  reach the apparatus with almost parallel
paths. Therefore, the distribution of their separation angles is a good
measurement of convolution of the scattering in the mountain 
overburden and the detector's intrinsic angular resolution.  This
space angle should be divided by a factor $\sqrt{2}$ to take into
account of the independent deviations of each muon in the pair. 
The reconstructed MPSF \cite{MACRO_MOON} is more peaked to small angular 
deviations, with respect to a normal distribution, but with long 
tails at large deviations.\\

In order to analyze both the Moon and the Sun shadowing effects, we have 
to take into account the finite size of these objects. Therefore we have
modified the true MPSF by selecting random positions on a disk having
the same angular dimensions of the Moon and then generating random
deviations according the MPSF. This new distribution, a convolution of
the original MPSF and an extended source of fixed angular radius, is
then used as the PSF for the analysis. The sensitivity of the results
to the angular radius of the object will be explored at the end of
this section.\\

A two dimensional likelihood analysis was performed, composing the
observed data with the expectation from a source of unknown strength
and position. We filled a two-dimensional histogram,
centered on the Moon position, and a similar histogram containing the
simulated events, obtained by adding to the expected 
background \cite{MUASTRO_PAP}, the events due to a source at $(x_s,y_s)$
having a chance strength $S_M$, spread according to the modified PSF. 
The source in this case should have negative strength, since we are
looking for an attenuation. Both the observed and simulated events are
plotted using horizontal coordinates (azimuth and altitude) and using equal 
solid angle bin ($\Delta \Omega = 0.125^\circ \times 0.125^\circ = 1.6 \times
10^{-2}$ deg$^{2}$). For each of the bin in the histogram used, we assume 
to have a source exactly at the bin center, then the unknown strength $S_M$ 
is evaluated by minimizing the quantity $\chi^2$ for poissonian
distributed data \cite{PDG}
\begin{equation}
\chi^2 (x_s,y_s,S_M)=2 \sum_{i=1}^{n_{bin}}[ N_i^{sim}- N_i
  +N_i\ln{\frac{N_i}{N_i^{sim}}}],
\end{equation}
\noindent
 where the sum is over all bins in the window \cite{PDG}, $N_i$ is the
 observed number of muons in the i-th bin, and $N_i^{sim}$ the number of
 events in the same bin from the simulated distribution. The minimum value found in each bin,   
 $\chi^2 (x_s,y_s,S_M)$ was then compared with 
$\chi^2(0)$ for the {\it null hypothesis} that no shadowing
source is present at the center of the bin ($S_M = 0$). We then fill
a new two-dimensional histogram filled with the  
quantity $ \lambda = \chi^2 (0) - \chi^2 (x_s,y_s,S_M)$.
 The most likely position of the Moon shadow is given
by the bin having the maximum value 
$\lambda^{max} \equiv \Lambda$.  Since there is only one free
parameter, the strength of the source 
$S_M$, $\lambda$ behaves like $\chi^2_1$, a $\chi^2$ distribution with
one degree of freedom \cite{COS-B}.
  The significance of the moon
detection is given by P($\chi^2_1\ge\chi^2_1(\Lambda)$).\\

\begin{figure}[t]
\begin{center}
\includegraphics[height=15cm,width=15cm]{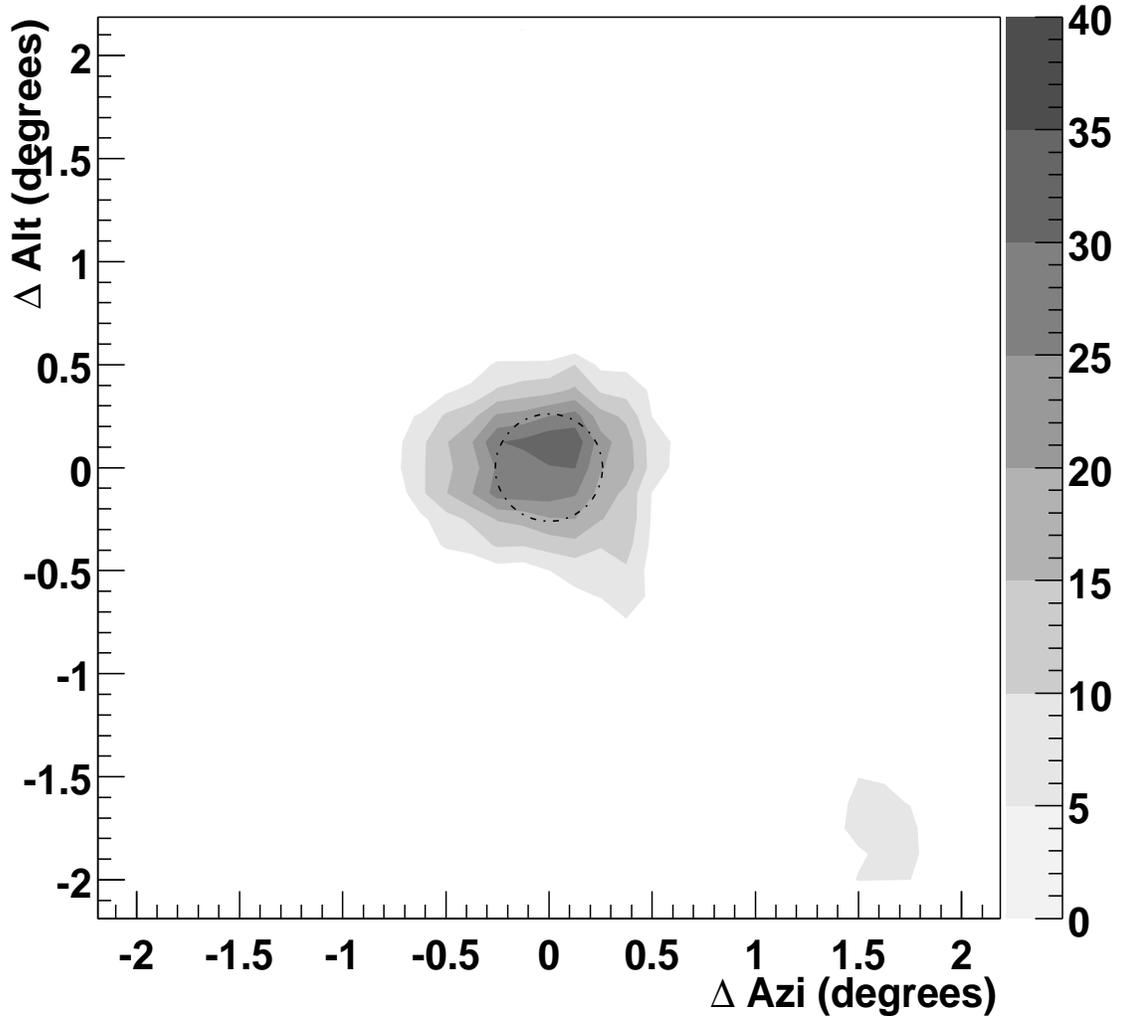}
\caption{The two dimensional distribution of $\chi^2$ in bins of equal solid
angle in the moon window. The axes are offsets from the moon center.
 A circle corresponding to the average lunar radius,
  $0.26^\circ$, is centered on the fiducial position of the moon, at
  position (0,0).
 The $\chi^2$ grey scale
  is given at the right margin of the figure.  The maximum of this
  distribution, $\chi^2 = 39.7$, is within the fiducial moon
  position at $\Delta \, {\rm Azimuth} = 0.\degr$, and $\Delta \, {\rm 
Altitude} = +0.125\degr$. The bin width is 0.25\degr.}
\label{moonlambda} 
\end{center}
\end{figure}

In Figure~\ref{moonlambda} we show the results of this analysis 
in a window $4.375^\circ \times
4.375^\circ$ centered on the moon position.  This window has been 
divided into $35 \times 35$ cells, each having dimensions 
$0.125^\circ \times 0.125^\circ$.
The $\lambda$ values of this figure are displayed in grey scale format
for every bin in the moon window. 
Also shown is the fiducial position of the Moon  
and a circle centered at this position corresponding to the average
lunar radius, $0.26^\circ$.
The largest deviation away from the expected background is found at 
 (0.\degr,+0.125\degr) with a  $\Lambda=39.1$, corresponding to a
 significance of 6.5$\sigma$ and 
 a negative strength of 316 events, as
expected for a shadowing effect. The $\chi^2$ value for the null
hypothesis, i.e. when a source of null intensity is set in the bin with
the largest deviation found, is 1240.71 for the $35\times35$ bins.
The entire sky region in the fiducial
position of the disk has $\lambda\ge 36.$ values. Moreover the
negative strength observed is equal to $316\pm 40$ events, in agreement
with the expected deficit. The $1\sigma$  error in the strength, treated as a
parameter, is directly estimated by the likelihood method by
considering the parameter interval delimited by the condition
$\lambda=\Lambda-1$\cite{PDG} (or $\chi^2=\chi^2_{min} +1$).\\

\begin{figure}[t]
\begin{center}
\includegraphics[height=15cm,width=15cm]{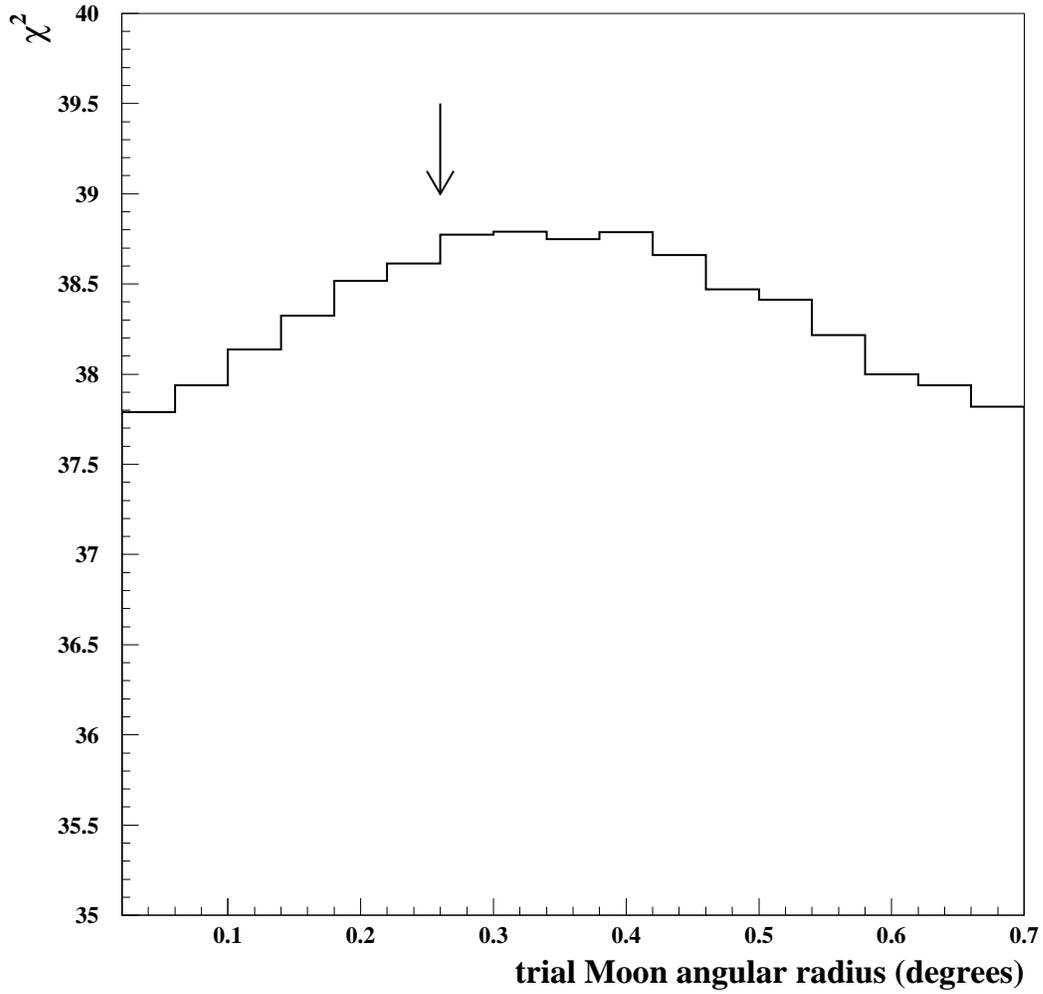}
\caption{Distribution of $\chi^2$ values at  the Moon shadow center
  versus the trial angular size for the Moon radius. The interval
  $\chi^2_{max}-1$ establishes the error interval for the Moon angular
  radius. The arrow shows the expected average value for the Moon radius. 
}
\label{moonradius} 
\end{center}
\end{figure}

We estimated other parameters used in the
simulation, like the Moon average radius \cite{casa}, in order to verify 
the correctness of the analysis or the MPSF shape itself \cite{MACRO_MOON}.
For this consideration  we assume that the true shadow center position
is  that having the largest deviation, i.e. at  (0.\degr,+0.125\degr),
and repeat the analysis using different PSFs each modified
by different values of the trial angular radius for the source,
starting with a null value corresponding to the true, unmodified MPSF.
This technique let us to both verify the sensibility of the analysis
to extended PSFs and also to quote an error on the radius of the
measured object.
 Fig.\ref{moonradius} shows the $\chi^2$ values at the
central shadow position, versus the trial angular disk radius. From
this figure we can quote as the radius for the Moon a value of
$(0.25\pm0.25)\degr$, well in agreement with the
correct averaged value. This result confirms that the Moon shadow
signal is more likely due to an extended source instead of a point-like
source.\\

The displacement of the shadow is in good agreement with the expected 
position, and this confirms the correct alignment of the
apparatus. We can quote as the maximum error in the alignment
of the apparatus a value of about 0.1$^\circ$, taking into account  the
observed  shadow displacement northward.

\section{Day-Night effects}
\begin{figure}[t]
\begin{center}
\includegraphics[height=8.6cm,width=8.6cm]{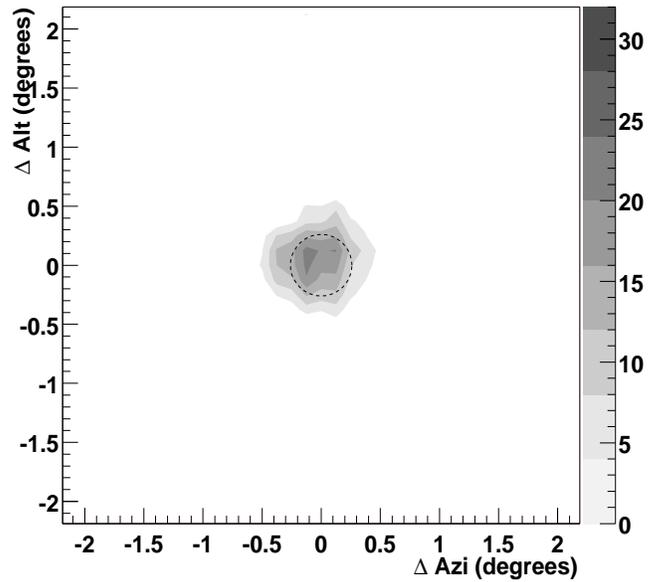}
\includegraphics[height=8.6cm,width=8.6cm]{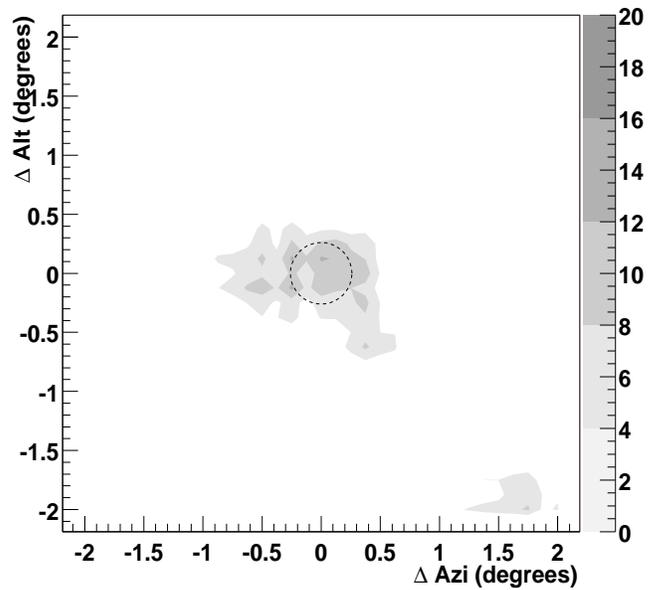}
\caption{The two dimensional distribution of $\chi^2$ in bins of equal solid
angle in the moon window for the ''night'' (a) and ''day'' (b)
subsamples. The axes are offsets from the moon center.
 A circle corresponding to the average lunar radius,
  $0.26^\circ$, is centered on the fiducial position of the moon, at
  position (0,0).
 The $\chi^2$ grey scale
  is given at the right margin of the figure.  
(a) The maximum of the
  distribution, $\chi^2 = 25$, is within the fiducial moon
  position at $\Delta \, {\rm Azimuth} = -0.1\degr$ and $\Delta \, {\rm 
Altitude} = +0.\degr$.
(b) The maximum of the
  distribution, $\chi^2 = 17.1$, 
  at $\Delta \, {\rm Azimuth} = -0.25\degr$ and $\Delta \, {\rm 
Altitude} = +0.\degr$. }

\label{moonday} 
\end{center}
\end{figure}

To study possible differences in the geomagnetic field due to the solar wind, 
we divided the muon sample in two subsamples by requiring that the angular 
distance between the Moon and the Sun is smaller or larger than 90$^\circ$. 
This requirement is almost equivalent to a daytime-nighttime requirement on 
the data, similar to what was used in a previous analysis \cite{tibet95}.
The results of the analyses for the two subsamples are shown in 
Fig.\ref{moonday}a-b: there is a sharper shadow for the ``night'' sample, 
with a large significance, and a broader shadow for the ``day'' sample 
(with a lower significance). Since the two subsamples are almost equivalent 
(198183 and 206805 events for ``day'' and ``night'' respectively) we conclude 
that night events encounter a reduced 
geomagnetic field with respect to the day events
\cite{Tsyganenko95,Tsyganenko96,Zhou}.
 The cause has probably to be ascribed to the
different configuration of the geomagnetic field in the two sides.
 Also time 
varying effects can be different for the two subsamples.  

\section{Shadowing effect of the Sun}
\begin{figure}[t]
\begin{center}
\includegraphics[height=15cm,width=15cm]{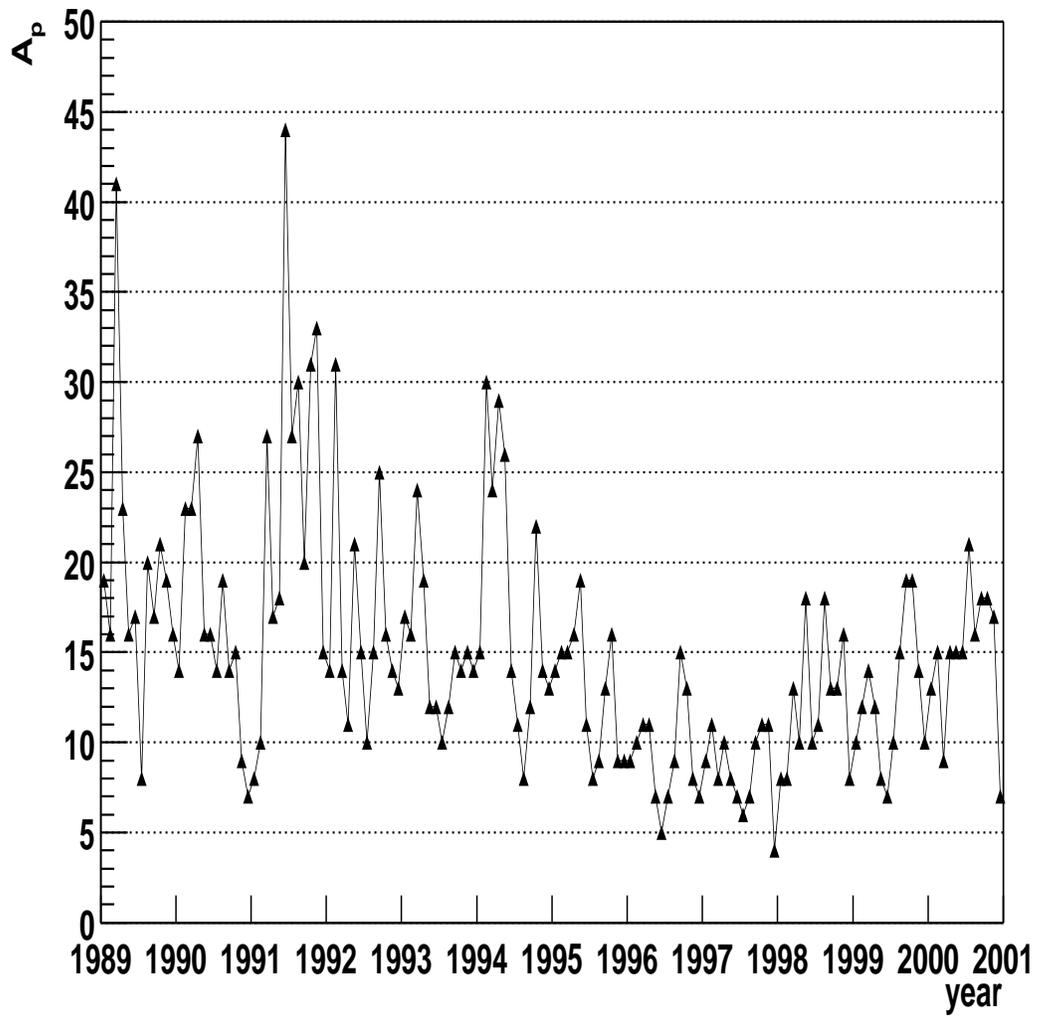}
\caption{Monthly A$\rm _p$ index in the 1989-2001 period.}
\label{Fig:Ap} 
\end{center}
\end{figure}

\begin{figure}[t]
\begin{center}
\includegraphics[height=15cm,width=15cm]{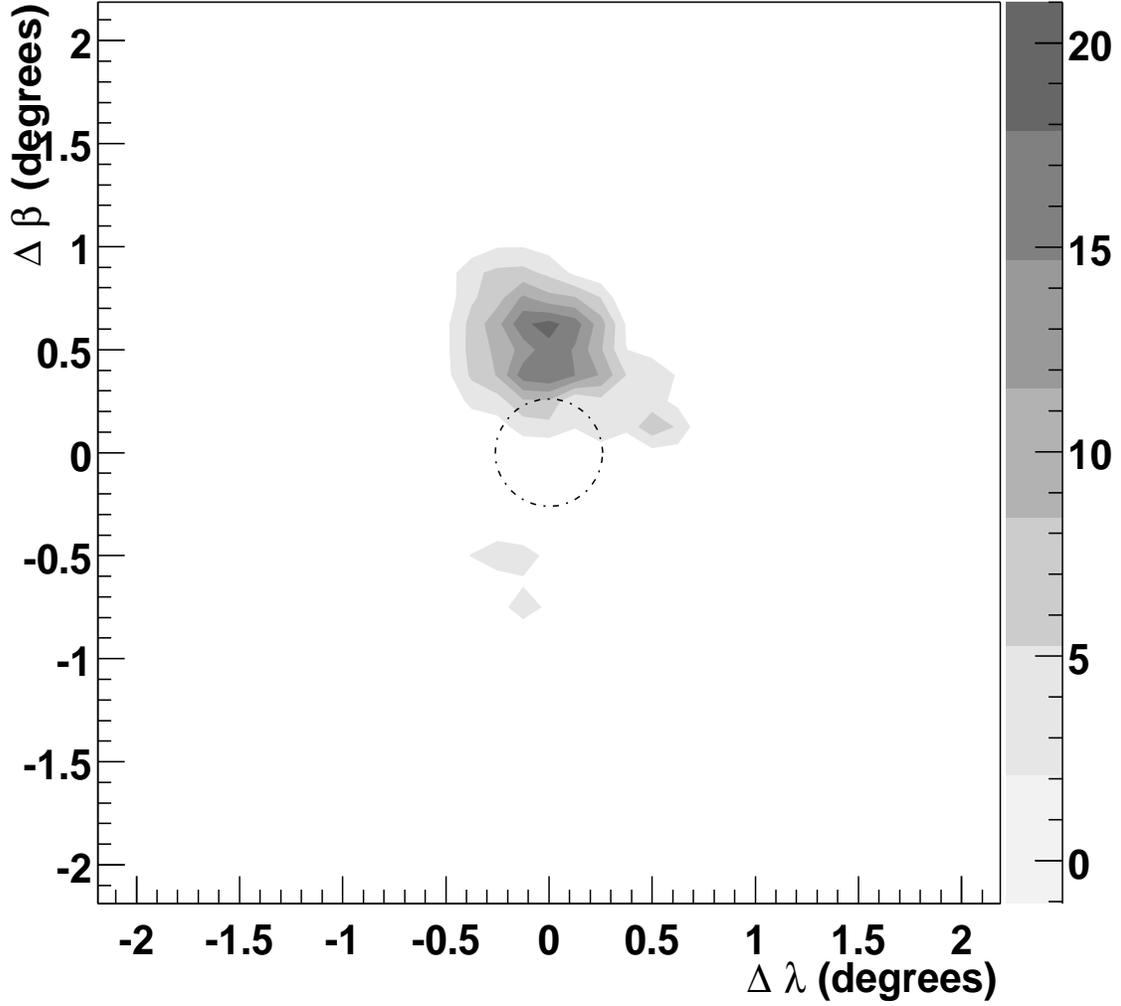}
\caption{The two dimensional distribution of $\chi^2$ in bins of equal solid
angle in the Sun window in ecliptic coordinates.
 The axes are offsets from the Sun center.
 A circle corresponding to the average sun radius,
  $0.26^\circ$, is centered on the fiducial position of the Sun, at
  position (0,0).  
 The $\chi^2$ grey scale
  is given at the right margin of the figure.  The maximum of this
  distribution, $\chi^2 = 22.0$, in the 
  position at $\Delta \, \lambda = 0\degr$ in longitude and $\Delta \,
  \beta = +0.625\degr$ ~in latitude. The bin width is 0.25\degr.}
\label{sunlambda} 
\end{center}
\end{figure}

The same analysis was performed also for the Sun. In this case the 
total number of events collected with an angular distance from the Sun
center of 10\degr is 396662, a sample almost
equivalent to that of the Moon. Since the angular dimensions of the two 
bodies are essentially the same, we expect about the same number of missing 
events. However, besides the geomagnetic field effect, two other
magnetic fields come into deflecting the cosmic ray particles:
 the Sun's magnetic field and the IMF. As these fields are variable in
 time, it is more difficult to predict the shadow displacement.
Using the ecliptic coordinates for the muons collected, and a window
centered on the calculated Sun position, we obtain the results shown in Fig.
 \ref{sunlambda}. The deficit is clear and it is displaced by 
0.6\degr northward with a $\chi^2=22.0$ 
corresponding to a signal significance of $4.6~\sigma$. The 
deficit observed for the Sun shadowing results to be $247\pm 48$
events, using the error estimated by the 
likelihood method as for the  Moon shadowing effect.
In trying to explain our result let us analyze the behavior of the
magnetic fields in the years during which we collected our data.
Fig. \ref{Fig:Ap} 
shows the monthly average A$_{\rm p}$ 
values \cite{Bartels} in the years 1989-2000.
These values indicate a global (planetary) magnetic activity and are
sensitive to solar particle effects on the Earth's magnetic field.
It is evident that the Sun was in a relatively quiet activity 
during 1993-2000, where the bulk of our events have been collected. 
Moreover, as we have mentioned in section 2.1,
 during the whole period the solar cycle was in the so-called
A$>$0 phase. In this situation the displacement due to the large
regular IMF is expected to be northward of the ecliptic plane. Let us
note that from 1991 till the end of 1997 the Heliospheric Current Sheet
was estimated to have a tilt angle $<$25\degr and was almost symmetric
between the north and south hemispheres of the solar cavity.

Since MACRO is located at a high 
latitude, it is possible that by averaging over a long period of observations 
the net result produces a displacement to the north as if the
 apparatus was mostly looking in the ``away'' hemisphere of the IMF.
The amplitude and the direction of the shift is in agreement with
other observations at the same energy range\cite{tibet96,tibet00}.

\section{Antiproton flux limits}

\begin{figure}[t]
\begin{center}
\includegraphics[height=15cm,width=15cm]{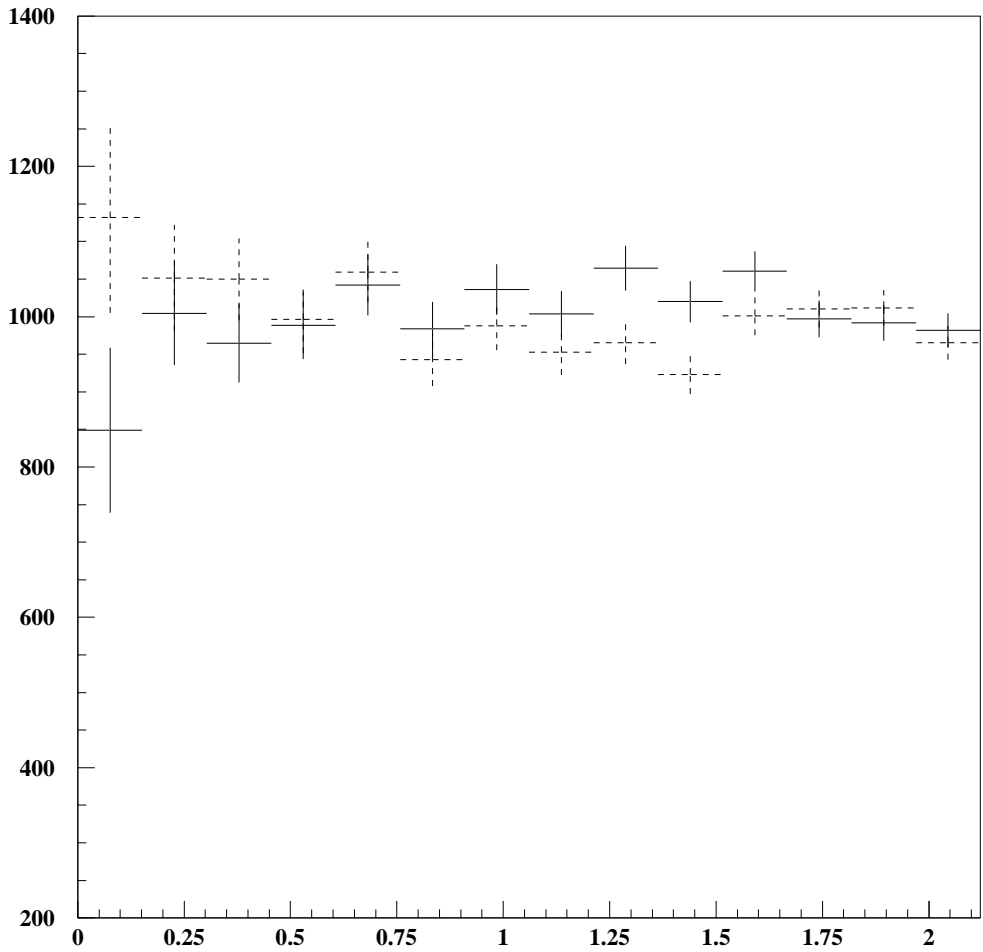}
\caption{Event density vs the angular distance from the Sun shadow
  position (0.6\degr northward with respect to the ``nominal
  center''). Superimposed the event density distribution centered in
  the symmetric position 0.6\degr southward.}
\label{sundensity} 
\end{center}
\end{figure}

The displacement of the Sun  shadow from the expected center position can be 
used to evaluate a limit on the antiproton abundance in the cosmic ray flux 
\cite{urban,chantell,artemis}. In fact if a primary antiproton flux is 
present, it should produce a displacement on the opposite direction with 
respect to that of the proton flux \cite{urban}.
The $\chi^2$ level in the symmetric position with respect to the sun
center, i.e. at  0.6\degr southward, results to be
$\chi^2=3.2$. In this case we use again  the likelihood analysis to
estimate an upper limit at the desired confidence level. Taking into
account that the $\lambda$ variable follows the $\chi^2$ distribution
with one degree of freedom,  we
can estimate the upper limit at 68$\%$ c.l. choosing 
$\Delta\chi^2=1$ and the upper limit at 90$\%$ c.l. choosing to $\Delta\chi^2=2.7$.  
Using this method we obtain $n_{68\%}=125$ events and
$n_{90\%}=155$ events of deficit. Therefore  $\rm  \overline p/p=51\%$ at 68$\%$
c.l. and   $\rm  \overline p/p=62\%$ at 90$\%$ c.l. for primaries 
at a mean energy of about 20 TeV \cite{tibet95}. 
We can also evaluate the upper limit looking to the event density
distributions centered in the position found for the sun shadow,
i.e. 0.6\degr displaced northward respect to the Sun ``nominal''
position, and in the symmetric position. The two distributions are
shown in Fig.\ref{sundensity}, and again is visible the deficit in
the distribution
northward position while the other is flat. Using the considerations
derived from   Fig.\ref{sigma_int} we found that the number of
observed events within 0.5$^\circ$, from the distribution southward
centered, gives 664 events observed and 710 expected. We can therefore
quote the upper limit as 60 events (68$\%$c.l.) and 81 events
(90$\%$c.l.).

Since as previously explained\cite{MUASTRO_PAP} the choice
of a 0.5\degr half-angle cone, collects
about 50$\%$ of events from a source, the upper limit should be $\rm \overline
p/p=120/247=48.5\%$ at 68$\%$c.l. and $\rm \overline
p/p=162/247=52.\%$ at 90$\%$c.l. in agreement with the likelihood estimations.

\begin{figure}[t]
\begin{center}
\includegraphics[height=13.5cm,width=13.5cm]{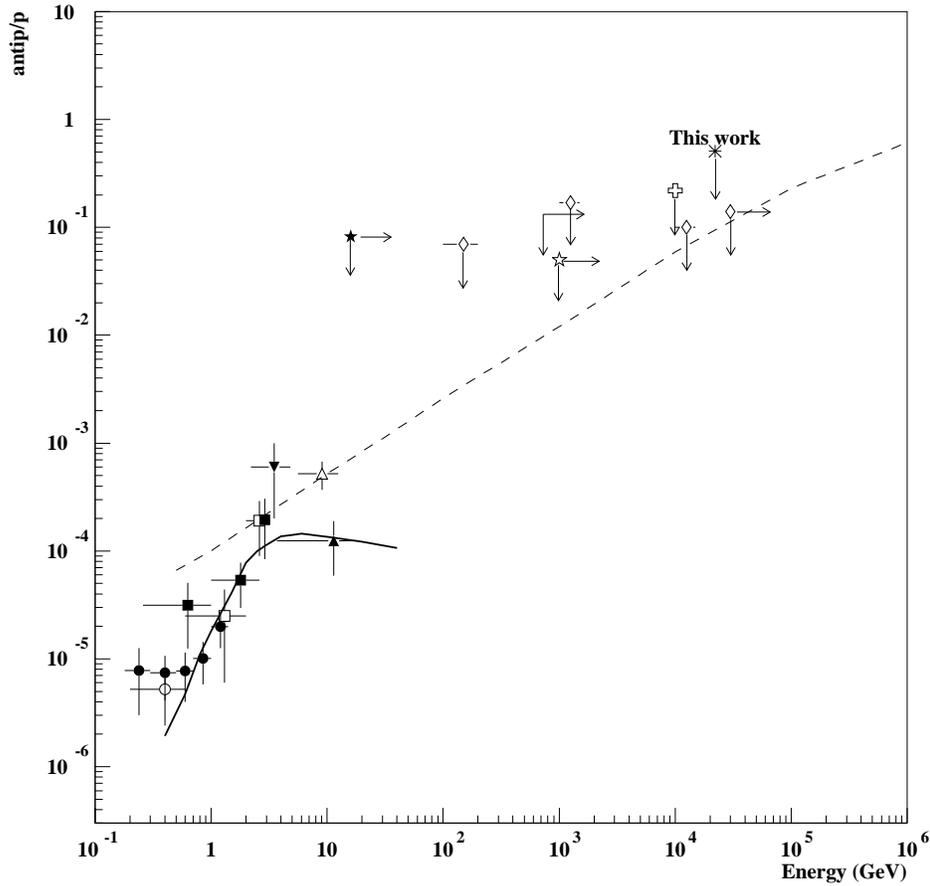}
\caption{\small Antiproton/proton ratio from various experiments at different 
average energies for period of A$>$0:
$\bigtriangleup$~R.L.~Golden et al., Phys. Rev Lett. 43(1979) 1196;
$\blacktriangledown$~E.A.~Bogomolov et al. Proc. 16th ICRC (Kyoto,1979) 
vol. 1,330;
$\blacktriangle$~M.~Hof et al. (MASS Collaboration) Astrophys. J., 467 (1996) L33;
$\blacksquare$~J.W.~Mitchell et al. (IMAX Collaboration) Phys. Rev. Lett. 87 
(1996) 3057;
$\ocircle$~K.~Yoshimura et al. (BESS Collaboration) Phys. Rev. Lett. 75 (1995);
3792 and  A.~Misseev et al. (BESS Collaboration) Astrophys. J. 474 (1997) 479;
$\square$~M.~Boezio et al. (CAPRICE Collaboration) Astrophys. J. 487 
(1997) 415;
\ding{108}~H.~Matsunaga et al. (BESS Collaboration) Phys. Rev . Lett. 81 
(1998) 4052;
\ding{80}~G.~Brooke and A.W. Wolfendale, Nature, 202 (1964) 480;
$\bigstar$~N.~Durgaprasad and P.K. Kunte, Nature, 234 (1971) 74;
$\Diamond$~S.A.~Stephens, Astron. Astrophys., 149 (1985) 1;
\ding{62}~\cite{tibet95};
no symbol~\cite{L3+CO};
$\ast$~MACRO this work. 
Solid line: unmodulated intestellar prediction. 
S.H. Geer and D.C. Kennedy, Astrphys. J. 532 (2000) 648
Dashed line: calculated antip/p ratio for extragalactic origin
S.A. Stephens and R.L. Golden, Space Sci. Rev. 46 (1987) 31
}
\label{Fig:antip} 
\end{center}
\end{figure}

Figure \ref{Fig:antip} 
shows our $\rm \overline p$/p limit compared to measurements from other 
experiments.

\section{Conclusions}
Using a sample of 50 million muons MACRO has detected the Moon and Sun 
shadows, with a significance of 6.5 and 4.6 $\sigma$, respectively. The 
strength of the signals is in agreement with those expected. 
The Moon shadow effect is centered as expected in the case of primary 
protons of 15 TeV energy (median), demonstrating the correctness and the 
stability of the detector pointing ability. The Sun shadow is shifted with 
respect to the ``nominal'' center position. The absence of a symmetric
shadow leads to an upper limit of the $\rm \overline p/p$ ratio of
48.5$\%$ at 68$\%$ c.l. for high energy cosmic rays. 
  MACRO is the deepest detector 
to observe the shadows of the Sun and of the Moon. This investigation confirms 
that the apparatus had the capability to  detect signals from cosmic sources 
by observing secondary muons underground\cite{MUASTRO_PAP,NIASTRO_PAP}.

\section{Acknowledgments}
We gratefully acknowledge the support of the director and of the staff of the 
Laboratori Nazionali del Gran Sasso and the invaluable assistance of the 
technical staff of the Institutions participating in the experiment. We thank 
the Istituto Nazionale di Fisica Nucleare (INFN), the U.S. Department of 
Energy and the U.S. National Science Foundation for their generous support 
of the MACRO experiment. We thank INFN, ICTP (Trieste), WorldLab and NATO 
for providing fellowships and grants (FAI) for non Italian citizens. 

\newpage

\newpage
\newpage

\newpage

\newpage

\newpage

\newpage

\newpage

\newpage
\thispagestyle{empty}

\end{document}